\documentstyle[12pt]{article}

\title{\bf Comment on ``Experimental Separation of Geometric and
Dynamical Phases Using Neutron Interferometry"}

\vspace{60mm}

\author{\bf Rajendra~Bhandari}

\date{ }

\begin{document}

\maketitle
\vspace{10mm}
\begin{center}
\begin{tabular}{ll}
            & Raman Research Institute, \\
            & Bangalore 560 080, India. \\
           
\end{tabular}
\end{center}
\vspace{40mm}
-----------------------------------------------------------------------\\
submitted to Phys. Rev. Lett.\\
MS version of 17 Oct 1997; Note and references added on 24 Feb 2001.

\newpage

In ref. \cite{wagh}, Wagh et.al. report a neutron interferometer
measurement of (a) a linearly increasing geometric phase shift
as a function of physical rotation $\delta\beta$ 
between two dual spin flippers in the range 
$-40^\circ$ and $40^\circ$ and (b) a $\pi$-phase shift
resulting from reversal of current in the two coils of the
flipper. While (b) is an interesting observation, we  
point out that extrapolation of the 
linear phase measured in (a) to larger values
of $\delta\beta$ and plotting the results of the two 
experiments, which are in different parameter spaces, 
on the same curve is unjustified, both theoretically
and  experimentally. Theoretically, the straight line 
in figure 2 of ref.\cite{wagh} 
ignores the factor $cos(\delta\beta/2)$ in the 
precession angle of the flipper, which is  intrinsic to 
the experiment and important for large  $\delta\beta$.
This  accounted for, the expected phase shift for large  $\delta\beta$ 
is neither linear nor purely geometric. It has a dynamical  term.
Experimentally, a continuously measured phase shift is defined in its
absolute value and not just  modulo $2\pi$. We show that
the equivalence of $\pi$-rotation of the flipper and
current reversal is only modulo $2\pi$.

Consider a version of the Wagh et. al.  experiment 
in which the currents in the two coils
of one of the flippers, producing magnetic fields 
$B_{1y}$ and $B_{2y}$ in the y-direction are  
varied independently while
those in the other flipper remain constant and the 
phase difference between the two beams is monitored at
small intervals in the parameter space spanned by 
$B_{1y}$ and $B_{2y}$, measured in units of the precession
angle in degrees. As long as the fringe contrast,
determined by the magnitude of the scalar product c of the 
spin states of the two
interfering beams is greater than zero, the phase difference,
defined by the phase of c (Pancharatnam criterion \cite{panch}), 
is measurable as  shift of the fringe maxima. 
The points in the $B_{1y}$, $B_{2y}$ plane where the 
two interfering spin  states  become orthogonal, form a lattice of 
phase singularities of strength +1 and -1. This happens
when the dual flipper produces precession equal to zero
or\\ $2n\pi$; $n$ being any integer. For example, the point
(0,0) has a -1 singularity and the point 
$q(180,180)$, where $q=(1/\sqrt2)$, has
a +1 singularity.  Field reversal corresponds to passage from 
the point I, i.e. $q(-180,180)$ to  F, i.e. $q(180,-180)$,
along some path in the parameter space.

Figure 1 shows the computed phase
evolution along three such paths  IAF (solid line),
IBF (dotted line) and  ICF (dashed line), where 
A, B, C are the points
$q(179,179)$, $q(181,181)$
and $q(1,1)$ respectively and IA, AF,
IB, BC, IC and CF are straight lines, each divided into 100 equal steps
for computation. 
One sees that the phase change for field reversal can be 
$+\pi$ or $-\pi$ depending upon 
the location of the path with respect to the 
singularities and can be highly nonlinear near a singularity. 
For closed paths, we find that the total phase change is 
$2\pi$ times the algebraic sum of the strengths of
the singularities. The choice of an appropriate path
can therefore yield, for field reversal, phase
change  $(2n+1)\pi$, where $n$ is any integer.

Finally we  draw attention to the fact that the 
effects reported in ref.\cite{wagh}, as
well as  the  singular effects described above have 
been seen in optical interference experiments using 
polarization states of light as the two-state system 
\cite{rbdirac,iwbs,rbreview}. 
For example, when circularly polarized
light passes through a pair of halfwave plates with
principal axes at $0^\circ$ and $45^\circ$, 
the $\pi$-phase shift observed when the first is
rotated through $45^\circ$ and the second through
$-45^\circ$   verifies Pauli
anticommutation \cite{anal}.  A recently proposed generalized
halfwave plate \cite{uhwp} can verify
all three Pauli anticommutations in a single 
experiment. \\
---------------------------------------------------
\newpage

Note Added:

Some of the landmark papers in the field, e.g. refs.\cite{panch,aa,jsrb},
while playing a key role in the development of the subject, define 
geometric phase as a quantity modulo $2\pi$, in the process missing the 
physical significance and measurability of the phase discontinuities and 
$2n\pi$ phases originating in singularities. This amounts to the 
proverbial ``throwing away the baby with the bath", a fact that becomes 
particularly obvious when dealing with systems with more than 2 states,
e.g. particles with spin $>1/2$. The weight of conventional wisdom 
inherent in the above formulations has prevented the appreciation of the 
fact (and publication of this comment) that $\pm 2\pi$ phase shifts 
resulting from circuits around singularities have been observed in 
interference experiments with polarized light and are observable in 
neutron interference experiments . 
It may be pointed out that Samuel and Bhandari \cite{jsrb} define the 
noncyclic geometric phase without the shortest geodesic rule for 
closing the open paths. The importance of this rule  
becomes obvious in ref.\cite{jumps} which first introduced the 
phase jumps in the context of the geometric phase. See also the 
discussion on page 1403 of ref.\cite{berrymo} which almost arrives 
at a $\pi$ - phase jump but suddenly becomes silent !

In the experiment of ref. \cite{wagh} analysed in this paper, the inadequacy 
of the ``phase modulo $2\pi$" becomes further transparent if one considered 
the same experiment with particles with spin angular momentum along the 
z-axis equal to $n\hbar/2$,  n being an integer $>1$. 
One would expect the measured geometric phase to follow a straight line 
beginning with a phase $-n\pi$ and ending at $n\pi$. With the approach of 
ref.\cite{wagh}, one would never get such a straight line for the measured 
phase.

\newpage

\section*{}{\Large {\bf Figure Captions}}

\addcontentsline{toc}{section}{Figure Captions}

{\bf Figure 1}:

The expected phase shift as a function of distance 
along three paths in the space of parameters $B_{1y}$, $B_{2y}$,
beginning at the point I, i.e. $q(-180,180)$
and ending at F, i.e. $q(180,-180)$ and passing
through the point $q(179,179)$ (solid line), the
point $q(181,181)$ (dotted line) or the point
$q(1,1)$ (dashed line). q stands for $1/\sqrt2$. 
Note the nonlinear variation and change in sign of the phase.

\end{document}